# Field tunable plasmonic lenses for optical microscopy

Divyansh Wadhwa,[1,†] Gurharinder Singh,[1,†] K. Balasubramanian [1,†]

[1] *Department of Material Science, IIT Delhi* [†]*These authors contributed equally.*

**Abstract:** We investigate the behavior and tunability of plasmonic lenses created from arrays of nanoslits, applicable in sub-wavelength optical microscopy and other high-resolution imaging systems. Simulations were performed on COMSOL Multiphysics to assess power flow and focal shifts in plasmonic lenses with differing slit designs, refractive indices, and angular distributions. The findings indicate that the confinement can be regulated by adjusting these parameters.

## 1 Introduction

In conventional optics, we can tune the focal length, by introducing spatially different optical path lengths (by means of multi lens setup, or introducing different mediums). Although this method is sufficient for bulk optics, it is not without any limitations, primarily because of the large size of the optical devices required to induce a sufficient difference in optical path lengths and the resulting high losses. Plasmonic lenses presents an alternative solution for achieving light focusing and beam shaping without the need to use bulk materials. Such lenses are referred to as metalenses, composed of patterned arrays.

Plasmon resonances can confine light to widths much smaller than the wavelength of light, leading to large near-field enhancements of the incident EM wave. This extreme confinement and enhancement properties provide unparalleled means for manipulation of light and its interaction with metals, well beyond the diffraction limit. And hence they have been looked into for potential uses in forming *Tunable lenses* or *Metalens* [1], [2], [3], [4], with tunable properties such as a variable focal length and resolution power, and very high resolution in general. Plasmonic lenses potentially can lead to lighter, cheaper and flexible lenses.

There are many designs both in experiments and simulations implementing such *lenses*. Lee et al [5] made and simulated one such implementation of a nanoslit through *nano slits* arranged in arrays with varying degree of rotations. We also find simulations of similar metalens designs on FEA software COMSOL Multiphysics®, such as done by Guo et al. [4] and Paasonen [6]. We aim to extend the simulation models, trying to focus on the tunability aspect.

## 2 Setup and Results

We first begin with a simpler singular nanoslit structure, and try to find confinement in it. We perform our simulations on COMSOL Multiphysics®. We use *Electric Wave Frequency Domain* (EWFD) module, where we use Maxwell's equations to solve the system. Then we consider a system as considered by Paasonen [6] and try to implement a similar system but with tilted slits. You can find the setup parameters and other approximations considered in the Supplementary . To get a sense of confinement, we consider Electric field confinement and the Power flux.

We consider a setup where we have a reflecting lens consisting of numerous nanoslits with distance between individual slits greater than SPP wavelength so that the plasmonic effect of each slit can be considered independent. We also assume that for most part the transfer of plasmon from metal to dielectric (where ever they make contact[1] ) are lossless.

Following factors (and how they can be tuned) affect the tunability of the focal length and the confinement :

---

[1]This was a problem of the setup considered, and discussed in more detail in

1. Variation of *wavelength* : This can be obtained by physically changing the light source i.e or by use of filters in incident light.
2. Variation of *refractive index* : Obtained by using materials with high electro-optical coefficients and changing the applied electric field across material.
3. Variation of *angular distribution*, *slit-width* and *slit-gap* : Using different sets of fabricated slits (having different curvatures) or by a global mechanical strain.

Another thing to note is that we can achieve variations through wavelength and refractive index and distribution can be varied independently of each other, and hence for the best case we can consider the effects to add up logarithmically. Changing the distribution is not easily achievable with a completely solid lithographic setup. We propose a setup where variable angular distribution can be obtained during operation.

Each nano slit causes phase retardation, and overall interference depends on this phase difference from the nano slits. The propagation constant $\beta$ can be found using the following equation. [3]

$$\tanh\left(\sqrt{\beta^2 - k_0^2 \varepsilon_0} \frac{w}{2}\right) = -\frac{\varepsilon_d \sqrt{\beta^2 - k_0^2 \varepsilon_m}}{\varepsilon_m \sqrt{\beta^2 - k_0^2 \varepsilon_d}} \tag{1}$$

Where $k_0$ is the wave vector in the free space, and $\varepsilon_d$ and $\varepsilon_m$ are the relative dielectric constant of the dielectric and metals respectively, and $w$ is the slit width.

And the total phase of the wave can be found using the following equation. [3]

$$\varphi = \varphi_0 + \Delta\varphi_1 + \Delta\varphi_2 + \beta d - \theta$$

$$\Delta\varphi_1 = \arg\left(\frac{n_1 - \frac{\beta}{k_0}}{n_1 + \frac{\beta}{k_0}}\right) \quad \Delta\varphi_2 = \arg\left(\frac{\frac{\beta}{k_0} - n_2}{\frac{\beta}{k_0} + n_2}\right) \quad \theta = \arg\left(1 - \left(\frac{1 - \frac{\beta}{k_0}}{1 + \frac{\beta}{k_0}}\right)^2 e^{i2\beta d}\right) \tag{2}$$

Where $\varphi$ is the total phase, $\varphi_0$ is the initial phase, $n_1$ and $n_2$ are the refractive indices of the 2 media before and after entering the slit, and $\Delta\varphi_1$ and $\Delta\varphi_2$ are the phase shifts of wave during entering and exiting the slit, $\theta$ is the phase shift considering reflections within the slit and $\beta d$ is the phase retardation caused by the slit.

For a parallel set of slits with varying widths [3], [6] the resultant phase distribution as a function of distance from origin, situated on axis of cylinder (if the lens is lying along the $yz$ plane, then along $x$ axis) is given by the following [3]

$$\varphi(x) = 2n\pi + \frac{2\pi f}{\lambda} - \frac{2\pi\sqrt{f^2 + x^2}}{\lambda} \tag{3}$$

To model a system with tilted slits, Lee et al [5] considered all the individual nanoslits to be individual dipoles. This can be considered if the wavelength of the formed SPP is small enough compared to the gap between 2 slits such that they dont interfere. Electric field is considered as result of tilting of the slit of the following form

$$E_z(x, y, z) = A(\theta(y)) e^{j\varphi(\theta(y))} e^{-\kappa_a z} e^{j(k_{\text{spp}}|x| - \omega t)} \tag{4}$$

where $k_{\text{spp}}$ is given by $k_{\text{spp}} = k_0 \sqrt{\frac{\varepsilon_1 \varepsilon_2}{\varepsilon_1 + \varepsilon_2}}$ is the wave vector for the SPP, and $\kappa_a$ is the dispersive term $\kappa_a = k_0 \sqrt{\frac{\varepsilon_2^2}{\varepsilon_1 + \varepsilon_2}}$. This is valid for $|y| < \frac{l}{2}$, where $l$ is the length (or the diameter) of the lens. They consider a parabolic distribution with a general form

$$\theta(y) = ky^2 \tag{5}$$

where theta is the angle of slit from the normal of plane Figure 1. We will also vary the $k$ parameter, which is the proprtionality constant, which can be controlled applying a some strain to the system which will be discussed later.

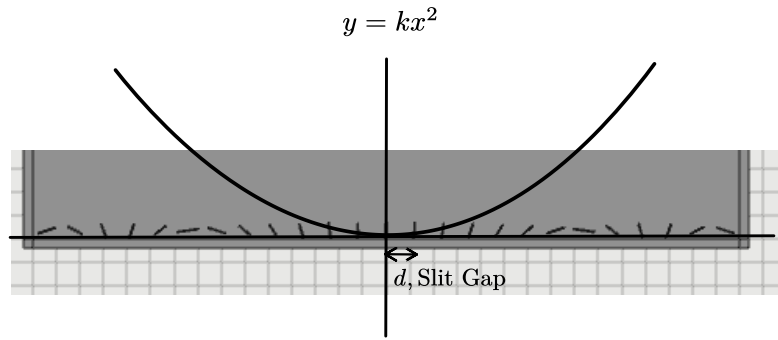

Figure 1: A schematic of the parabolic distribution of tilted nanoslits, for $k = 2$ and $d = 600$ nm

We tried to replicate the results from Lee et al [5] and found values of $\lambda = 275$ nm $\quad n = 2.1 \quad k = 1 \quad d = 600$ nm $\quad h = 350$ nm, where $d$ is the gap between midpoints of the slits (which remains unmoved in rotation), and $h$ to be height of initial slit[2] to match very closely with their results Figure 2. These values are not a unique combination, there are many possible sets.

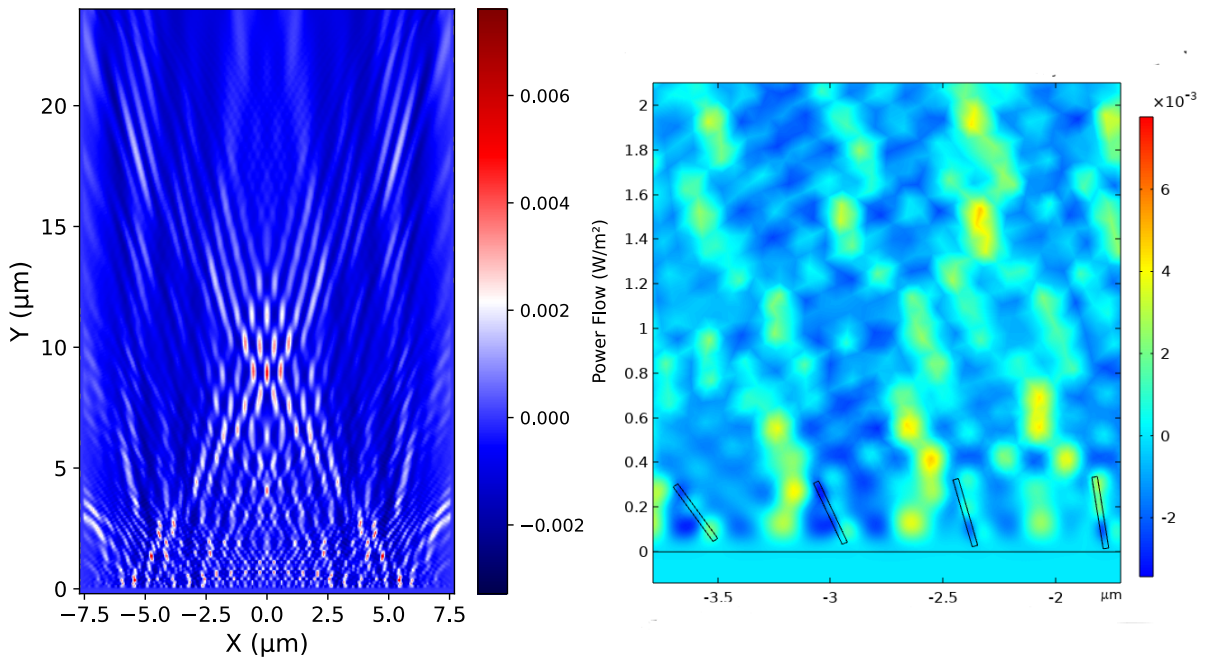

Figure 2: (a) Power flux plot showing convergence of a parallel beam incident on a system with $\lambda = 275$ nm $n = 2.1 k = 1 d = 600$ nm $h = 350$ nm. The observed focal length in this case is about $8.85 \mu m$ (b) A close-up picture of slits far from origin, which are detached, where flux does not get concentrated in the slit, and hence, such slits are not effectively interfering

Our simulations are generally more noisy (ratio of noise to actual power flux at focus) than Lee et al [5] or Paasonen [6]. Our current hypothesis for this is the limitation of how we generate our nanoslits,

---

[2]While this should remain the same, due to limitations mentioned previously, they also change with angle

and how there are places where there is not enough contact. Figure 2 This only happens here because the slits which end up having a larger gap from the surface end up having a different (lower) amount of flux as compared to ones which are at a lower angle. This resulted in a messier flux distribution.

Here is a side by side comparison of shift of focal length with change of wavelength.

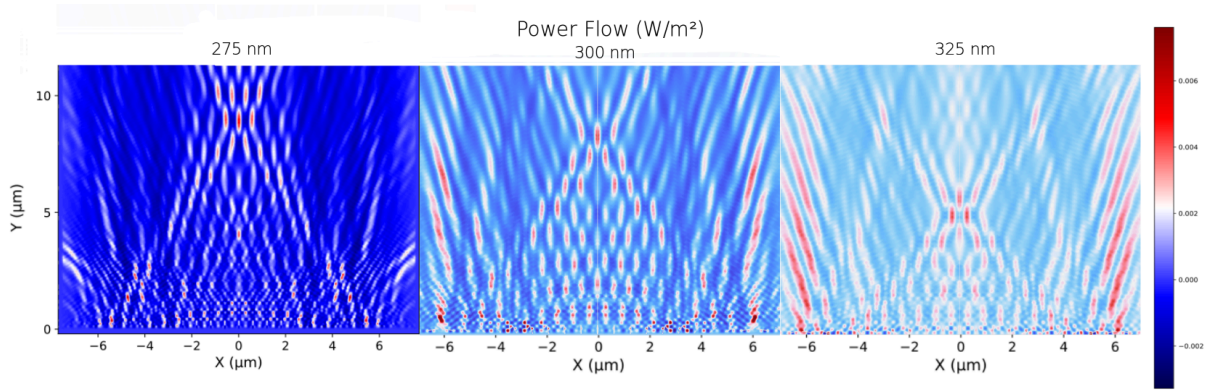

Figure 3: Reduction of focal length as we change wavelength from 275, 300, and 325 nm

Here is a plot to show variation of focal length by varying wavelength, k, and slit gap.

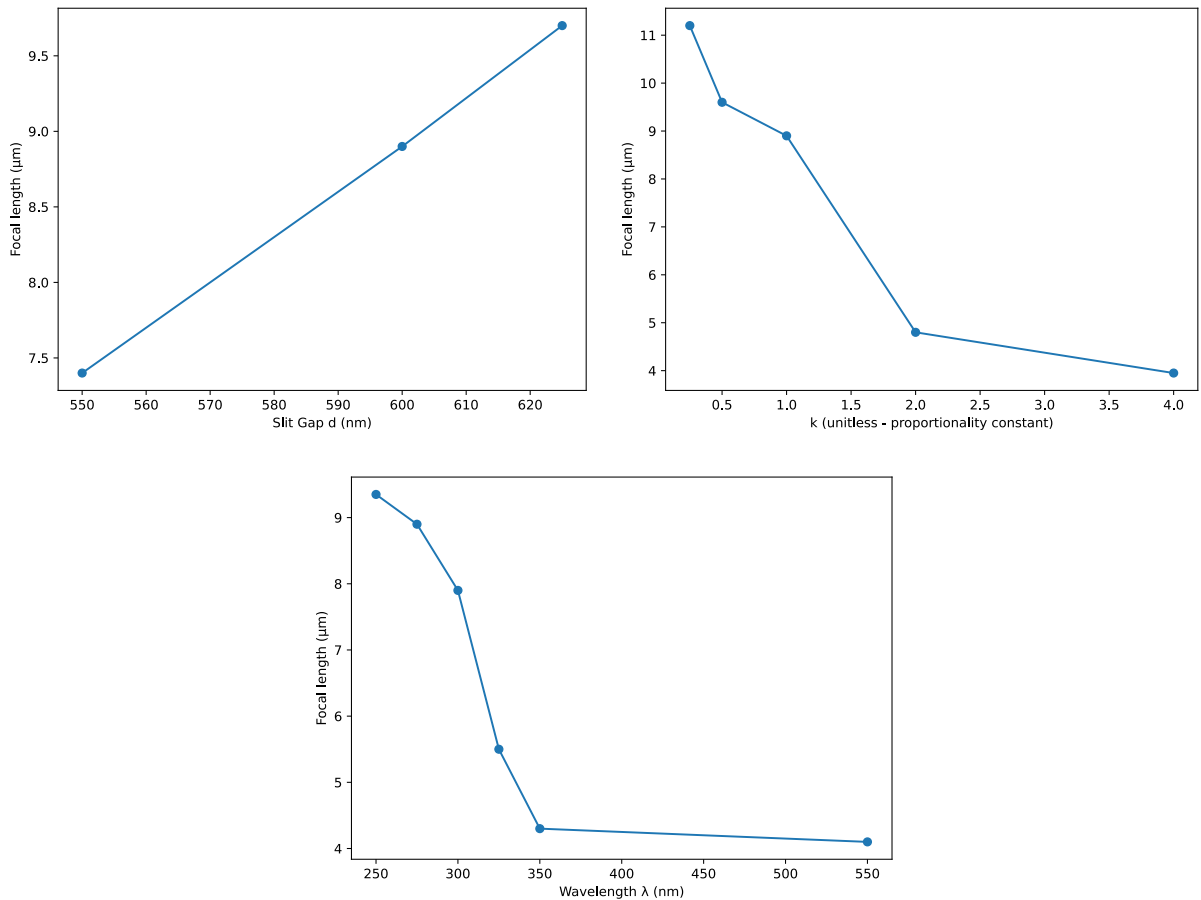

Figure 4: Variation plots of focal lengths vs (a) Slit Gap $d$ (b) Proportionality constant $k$ (c) Wavelength $\lambda$ $n$

Change in focal length due to wavelength $\lambda$ also depicts the chromatic aberration due to this lens. We tested for values with a step size of 25 *nm*, and 1 for double the wavelength. As the wavelength increases, we generally see a downwards shift of focal length, but we start seeing a plateau, as beyond certain wavelength, the noise is too high, and we do not observe a clean focus[3]. $k$ can not be changed while operation, and essentially is defined by initial design in a lithographic design. If we consider flexible slits, this can also be changed. $d$ can be changed by applying a global strain. We have considered a situation of $\pm 10\%$ strain. Variation due to change in $n$ of material due to electro-optic coefficient, in best cases can be of order of 0.1, and we consider that range around initial values, but change in focal length is not very significant (not included in figure).

## 3   Discussion and Future work

The results demonstrate that in principle we can achieve a tunability in the focal length of the plasmonic lens. This may just involve a small amount of mechanical strain, which can be applied to a polymeric bet material holding our setup, which can be finely applied with something as simple as a screw. Real problem remains in the creation of such tilted slits. While these are possible by electron lithography, they are limited by cost and scalability, and availability of suitable materials. We can also possibly look forward to suspended CNT on a bed. These CNTs can be carboxylated [7] and can act as a dielectric material. Under suitable environments (for example in a mildly alkaline solution), these can form carboxylates, and hence become charged. If we suspend the whole setup under an alkaline viscous fluid, we can have our nanoslits essentially arranged, but erect. On application of a varying electric field, these can bend (following the field lines) and can possibly be arranged in a fashion as discussed above.

---

[3] When focal length approaches the $k_{\text{spp}}$, we no longer see clear interference pattern, and start observing a *YDSE* like pattern. we specifically tried for 550 *nm*, since it is double that of 275 *nm*, and we observe intense YDSE like alternating high and low intensity spots up until 4-5 $\mu m$, but we also observe a much lighter focus, centered around 8.85 $\mu m$, which maybe formed due to higher order terms